# An Efficient Framework for Network Code based Multimedia Content Distribution in a Hybrid P2P Network

M. Anandaraj, K. Selvaraj, P. Ganeshkumar, K. P. Vijayakumar

[1]*Associate Professor*, [2,3]*Professor*
*Department of Information Technology,*
*PSNA College of Engineering and Technology, Dindigul, Tamilnadu*
ananddgl@yahoo.com, selme@gmail.com, msg2ganeshkumar@gmail.com

[4]*Assistant Professor*
School of *Computer Science and Engineering,*
*SRM institute of Science and technology, Chennai, Tamilnadu*

kalkivijay@rediffmail.com

**Abstract**     Most of the existing P2P content distribution schemes carry out a random or rarest piece first content dissemination procedure to avoid duplicate transmission of the same pieces of data and rare pieces of data occurring in the network. This problem is solved using peer-to-peer (P2P) content distribution based on network coding scheme. Network coding scheme uses random linear combination of coded pieces. Hence the above stated problem is solved ease and simple. Our proposed mechanism uses network coding mechanism in which several contents of same message is grouped into different group and coding operation is performed only within the same group. The interested peers are also divided into several groups with each group have the responsibility to spread one set of contents of message. The coding system is designed to assure the property that any subset of the messages can be utilized to decode the original content as long as the size of the subset is suitably large. To meet this condition, dynamic smart network coding scheme is defined which assures the preferred property, then peers are connected in the same group to send the corresponding message, and connect peers in different groups to disseminate messages for carrying out decoding operation. Moreover, the proposed system is readily expanded to support topology change to get better system performance further in terms of reliability, link stress and throughput. The simulation results prove that the proposed system can attain 20–25% higher throughput than existing systems, good reliability, link failure and robustness to peer churn.

## 1. Introduction:

Distribution of content over the internet refers to the delivery of digital content such as text and multimedia content, software and live streaming audio and video to a large group of users. Traditionally it is done by using the client-server system in which client sends request to the server and receives requested file as a response from the server[1]. To improve the performance of the system with respect to response and process time, a content delivery network(CDN) has been introduced. It locates a collection of servers in geographically distributed way. Commercial CDNs such as Akamai and Digital Island offer this service for many popular commercial sites. Peer-to-Peer content distribution model (P2PCM) has established itself to be an efficient, scalable, and a robust networking application to offer services for sharing large scale content and personal communications [2]. Instead of the traditional client-server model, peer-to-peer network merges the resource from all peers together in the network and contribute to all peers in return, which is the reason of its success [3]. P2PCM is a blend of a distributed storage function, a dissemination function of content among peers in the network, and a demand routing function. It is a distributed and cooperative architecture. In a P2P network model, each peer receiving help from other peers in the network for further distribution of content to others. These networks are scalable one. The new nodes join the network not only bring new demands, but also give additionally more bandwidth and computing power to the network. There is no longer a single point of crash in the network. Even a P2P network has the adequate network resources, it still faces confront of how to utilize those resources optimally without having any central coordinator as well with low complexity [4]. Finding an ideal route for packets is a difficult problem in the large scale content delivery network. These networks are highly dynamic such that nodes joining and leaving the network at any time and node have only local topological information about the network[5].

P2P content distribution schemes [2], such as Gnutella, KaZaA, Napster, and BitTorrent, have been broadly adopted to quickly distribute content over the Internet. P2P model has been installed over the Internet recently to offer on-demand or live media streaming services. Examples of P2P streaming schemes [6,7] such as TVants, CoolStreaming, PPLive, and PPStream. Either it is P2P content distribution or video streaming, a well-designed P2P scheme must contain a close to best possible rate of content disseminate and low complex mechanism for execution. Most of existing P2P schemes implements routing [3], such as store and forward, as the delivery model. BitTorrent is a popularly known P2P system among others. It uses swarming techniques to simultaneous download of a different segment of same content among peers. Downloading content by collecting its segment can be to extent modeled by typical coupon collector problem. In this, probability of gathering fragment comes down rapidly with the number of those already received. As the number of nodes increases, it is difficult to schedule the distribution of packet to other nodes optimally. Download local rarest segment first is a possible solution for the above said problem. But, most often such segment fails to match those that are globally rarest one. This affects other peers such as slower download rate and increase of failure rate.

In this paper, it is aimed to go ahead of the simple XOR to more general encoding operations and realize a better coding gain in dynamic P2P network. For clustered P2P networks, new coding based schemes to conduct packet coding with more general coding operations instead of simple XOR operation is proposed, which overcome the above limitations of the available coding-based schemes. In summary, the major contributions of this work are given as follows:

- The use of network coding using coupon collector problem is analyzed which is directly related to the difficulty of content distribution over P2P networks. BitTorrent network is considered for the analysis and explored the opportunity to use network coding approach in content distribution.
- A new network coding framework is developed, called as dynamic smart network coding (DSNC) for P2P content distribution in which peers within the campus network is grouped based on the similarity interest with different content of the same file. Hence the traffic in access link is reduced.

The rest of the article is prepared as follows: the literature survey is given in section II and the operation of network coding is described in Section III. The use of network coding mechanisms using coupon collector problem is described in Section IV. The proposed systems specification and technique are explained in Section V. The experimental setup for proposed system and performance evaluation are described in Section VI. Finally, the conclusion and future scope of this work is mentioned in Section VII.

## 2. Related Work

Network coding has been employed in several networks such as Storage Area Network, Peer-to-Peer networks, traditional wireless networks, video multicast networks, wireless sensor networks and several others. It has been proved that, with network coding, the cut-set bound of information flow rates in both P2P and multicast communication can be attained[8]. The core idea of network coding is a paradigm shift to let encoding at intermediate network nodes between the source and destination in a communication session. With the ability to code at intermediate nodes, there is a possibility to store, code and forward incoming packets in the communication session. This capability is in contrast to traditional commodity flows, where intermediate node only forwards the incoming flow. P2P network is the ideal place to implement network coding mechanism because of the random creation of P2P network topology, simple and easy to modify the topology and the nodes in a P2P network are end hosts which can carry out more difficult operations such as decoding and encoding operations.

Microsoft's Avalanche [9] is a model of P2P scheme attempting to solve piece and peer selection problem using network coding techniques, introduced by Gkantsidis and Rodriguez. It initiated several research interests to use network coding for distribution of large scale multimedia content. Instead of transmitting the original file segments, peers generate linear mixture of the segment they already have in its buffer. These combinations will be sent along with coding coefficients in each combination. After receiving sufficient linearly independent mixture of the original segment, peer can decode and construct the original file. This technique makes each packet equally important. Network coding offers optimal performance while network operating in a decentralized or distributed fashion. This attracts application in dynamically changing environments [10]. If the network makes use of the network coding technique, then nodes in the network can able to operate in a decentralized fashion without using overall topological knowledge of the network. Various tests and measurements were conducted out on the P2P network, and results together with the number of observations shows the advantages of using network coding in P2P network system. Applying network coding in P2P file sharing has two advantages such as content sharing becomes robust against peer departures and content propagation scheme simpler. Fine-tuning this scheme for particular applications, such as Video-on-Demand and file sharing are needed. The theoretical study that support the experimental evidence related with advantages of network coding, is missing[11].

In the last decade, there were several research works on the linear network coding in the literature. Li, et al. [12] demonstrated that linear network coding over a finite field is enough to attain multicast capacity. Kotter, et al. described an algebraic characterization for a linear network coding system in [13]. They also demonstrated an upper bound on the field size and a polynomial time algorithm to verify the soundness of a network coding method. In [14] [15], Lun, et al. introduced a distributed algorithm to identify a sub graph of the original topology such that the connection cost can be reduced without sacrificing multicast capacity. Ho, et al. described a random linear network coding (RLNC) approach [16] in which nodes produces edge vectors at random. The linear network coding scheme generated by this approach is not suitable at all the times. They showed that the probability of failure is the size of the finite field. According to [17], using RLNC pieces can lead to better scattering of pieces throughout the entire network. It reduces the possibility of nodes downloading the same pieces from others in the network. However, the more number of pieces of large content may decrease the probability of the occurrence of linearly dependent coded data. This technique significantly increases the need for decoding the content more often. To reduce the computation overhead, larger files are divided into a number of segments [18]. Each segment acts as an independent generation in network coding [19]. In contrast to the RLNC, Jaggi, et al. introduced a polynomial deterministic algorithm in [20] that can make deterministic linear network coding system for multicast networks. Min et al. demonstrated a scheme to use network coding for peer-to-peer file distribution which utilizes a P2P network to disseminate files located in a web server or a file server. A special type of network topology called as combination network, is used in their scheme. It was confirmed that combination networks attain unbounded network coding gain calculated by the ratio of network throughput with network coding to that without network coding [8].

The simple operation (XOR operation) to combine the packets in the network code based distribution has the benefit of encoding and decoding the packets faster. Even though encoding packets with the simple process over $GF(2^q)$ has merits, there are two main restrictions. First, only the undelivered native packets with distinct intended peers can be encoded together. Hence the possible coding opportunities cannot be utilized fully. Actually, the undelivered native messages with common intended peers also have the potential to be coded together with more common coding operations for improving transmission efficiency. Second, the search for the best possible set of undelivered packets to code is complex. It is a NP-complete problem. The following problems are indentified in RLNC, based on extensive literature survey:

- Computational complexity of decoding operation is very high because of Gauss-Jordan elimination scheme
- Transmission overhead is high due to the inclusion of large coefficients vectors in each coded packet
- The number of innovative information flow is reduced because of linear dependency among coefficient vectors of coded packet

There is an attempt to solve the second and third issue, made in this article by proposing new framework.

## 3. Network Coding background

In this section, the basic principle and some significant advantages of network coding are briefly described. Network coding has been proved to improve the performance of communication networks in both wired and wireless network in terms of throughput, robustness, security, etc [21 - 23]. It gives a computationally tractable solution for NP-hard problem that is computationally challenging in case of using traditional routing. It is useful mechanism to find a solution which achieves the optimal system performance. For example, in a P2P content distribution network, finding a best possible approach for routing is challenging due to large system size. It makes tough to trace data identities of all messages. If network coding technique is used, randomization of random linear combinations of packets can reduce this complexity since tracking packet uniqueness is not required. Only the quantity of data is substance.

Random linear network coding was first introduced in information theory [21] and it has been used in P2P content distribution to improve system performance in terms of download time and throughput [22][23]. Each peer can able to generate linearly coded blocks using the block it receives from its different incoming link and transmits the generated coded blocks to its downstream peers in network coding based P2P network. This technique simplifies P2P protocol design by making each packet equally important. Hence, data scheduling is easier. Let us consider two P2P content distribution, as demonstrated in Fig. 1(a) and Fig. 1(b). The first one don't make use of encoding at all, while the second one uses network coding across all existing received blocks. It is much more difficult for a non-coding protocol to guarantee a uniform delivery of all blocks in dynamic P2P networks. In Fig. 1, the source S has to send two messages to all three nodes in the bottom of the tree. It sends m1 and m2 via the link between S-P and S-Q. If the node S doesn't perform the coding operation, it should take a decision on which packet needs to be sent over the link between S and R. Node S has two a choices, either m1 or m2 send over the link which leads to inefficient use of available bandwidth. It is represented in Fig. 1 a. If network coding is performed in node S, the node S will send linear combination of m1 and m2 through the link between S and R. Hence all the three nodes in the bottom of the tree can decode and obtain those two packets in the efficient way. It leads to maximize the utilization of available resources in the network.

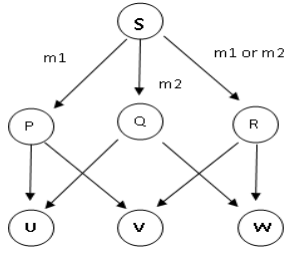 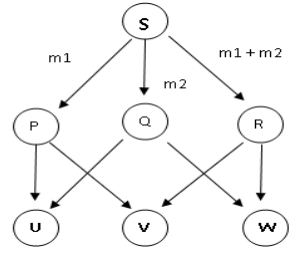

Fig. 1 a.                           Fig. 1 b
Fig. 1. Content Distribution with and without network coding

The above example reveals the advantages of network coding with respect to throughput also. It shows that network coding improve the throughput of multicast. In the above network topology, a server or source node S and a set of destination nodes D = {$d_1, d_2,...$}, multicast capacity is described as the maximum rate at which the server able to transmit the messages to the set of destinations, that is, equivalent to $\min_{d_l \in D}(Maxflow(S, D_i))$ where Maxflow(S,D$_l$) is the highest flow between source S and set of destinations D. In Fig. 1, multicast capability of the network is equal to 3. Without use of network coding, throughput is equal to 1.5, whereas with network coding the throughput is equal to 3. In general, Ahlswede et al. [20] has confirmed that if network coding is performed in network nodes, then the maximum multicast capacity can be obtained. Hence network capability can be fully used in multicast.

The network coding is used to reduce the download time because peers will not be further delayed to locate rarest blocks in the network. The problem of finding rarest blocks in a non-coding protocol does not exist in a network code based approach because of all coded blocks is equally innovative and useful to other peers in the network. However, network coding may not recognize its advantages without introducing considerable computational complexity. In this paper, group formation among peers and dynamic network coding mechanism are proposed, both aim to diminish the computational complexity of network coding.

## 4. Coupon collection problem and Linear Combinations

In this section, the coupons collector problem is described which is directly related to the problem of content distribution over P2P networks like BitTorrent [24]. Here coupon is related to the chunks in P2P network. Each peer completes its download after collecting all the chunks of original content in the network. In the coupon collector problem, a coupon is generated from a set of size S uniformly at random. The copy of token is located inside a container of some service. An entire version of containers is created in this way and placed on the market. A collector used to buy containers in order to collect the S coupons. He may have a restricted budget and also limited time to spend for gathering his collection. Thus, he needs to know the following:
1. The sample size $N_i$ must be known to acquire i distinct coupons and the sample size essential for obtaining all S coupons.
2. The waiting time $W_t$ to obtain the i[th] distinct coupons.
3. The number of different coupons $D_s$ such as collection size in a sample of size n. This value is of particular attention when the collector's resources such as time and money are limited and he can obtain only so many coupon containing containers.

In the above $N_i$, $W_t$ and $D_s$ are random variables. Their probability distributions are computed. The process of collecting the i[th] coupon in random is Bernoulli process. In each tryout the probability of success, such as drawing the i[th] distinct coupon is

$$P_i = 1 - \frac{i-1}{s} \geq \frac{1}{s}, 1 \leq i \leq s \qquad (1)$$

If more number of coupons collected already, then smaller the probability of success. To acquire the i[th] coupons, the collector has to craft on the average

$$E[W_t] = \Sigma_{j=1}^{\infty} l, p_i (1-p_i)^{j-1} = \frac{1}{p_i} = \frac{s}{s-i+1} \quad draws \qquad (2)$$

Here, also it is noted that the larger the number of previously collected coupons the longer time to wait to collect the new one. The sample size of collecting the i[th] coupon is $N_i = W_1 + W_2 + ... + W_t$, if the number of attempts made to collect the $i^{th}$ coupon is $W_t$. Hence the average sample size needed to collect the i different coupons is

$$E[N_i] = E[W_1 + W_2 + \ldots + W_R]$$
$$= s\left(\frac{1}{s} + \frac{1}{s-1} + \ldots + \frac{1}{s-r+1}\right) >\approx s\log\frac{s+1}{s-r+1} \qquad (3)$$

Here, bound is the left Riemann sum of 1/s on the interval [S – r, S]. It becomes tense for large value of S. Hence the average waiting time to collect S different coupons is bounded as follows

$$E[N_i] \geq S\log(S+1) \rightarrow E[N_i] = S\log S + O(S) \qquad (4)$$

If a collector has some constraints which will allow him to collect a sample of at most N coupons, he wishes to know how many different coupons $D_s$, he has in this sample. To calculate the expected value of $D_s$, we define the probability of not having a particular coupon in a sample of size n is $[(S – 1)/S]^n$. We can obtain the expected value of Ds as follows

$$E[D_s] = S[1 – (1 – 1/S)^n] \geq S(1 – e^{-n/S}) \qquad (5)$$

Now let us consider the collector need to collect not physical entities, instead he has to acquire information that can be processed in a way that can't be done on physical entities. Suppose the goal is to collect S numbers $p_1,\ldots,p_s$ which are elements of some Galois Filed GF(q). There are two different ways for the collector to acquire the numbers. The first method is to collect a number with replacement, which is similar to the classical coupon collector problem. In the second method, as a result of each draw a linear combination over GF(q) of the numbers is made and then coefficients and the result of combining are informed to the collector. Thus, with each draw, the collector collects a linear equation that unknown values of p have to satisfy. The collector has to collect a set of S linearly independent equations. The vector (coupons) said to be innovative if the draw results in equation coefficients are linearly independent of those already collected. Here innovative referred to the distinct coupons in the classical coupon collector problem. The probability distributions of the two random variables $N_i^c$ and $W_i^c$ are calculated, c represents coding used.

The distribution for the probable waiting time $W_i$ to collect the $i^{th}$ innovative coupon are calculated first. Once i-1 innovative coupons have been received, the random process of collecting the $i^{th}$ equation is a Bernoulli process. In each attempt, any of the $q^s$ vectors in $GF(q^s)$ can be drawn. The attempt is successful if the received vector is not one of the $q^{i-1}$ that belongs to the (i-1) dimensional space spanned by i-1 already collected innovative vectors. Therefore,

$$p_i^c = 1 - \frac{q^{i-1}}{q^s} \geq 1 - \frac{1}{q}, 1 \leq i \leq s \qquad (6)$$

If the collector already receives the larger number of innovative coupons, then the smaller the probability of success, there will be a nontrivial lower bound on this probability independent of n which can be made randomly close to 1 by increasing the field size q. To collect the $i^{th}$ innovative coupon, the collector has to make on average

$$E[W_i^c] = \sum_{i=1}^{\infty} m.p_i(1-p_i)^{m-1} = \frac{1}{p_i} = \frac{q^s}{q^s - q^{i-1}} < \frac{q}{q-1} \qquad (7)$$

attempt. The average sample size for the collection of i distinct coupons are calculated now as follows

$$E[N_i^c] = E[W_1 + W_2 + \ldots + W_i]$$
$$= \frac{q^s}{q^s} + \frac{q^s}{q^s - q} + \ldots + \frac{q^s}{q^s - q^{i-1}}, \qquad (8)$$

and this value is bound as follows

$$i < E[N_i^c] < i \cdot \frac{q^s}{q^s - 1}$$

Thus the average waiting time to collect S innovative coupons is

$$E[N_i^c] \cong s \text{ as } q \rightarrow \infty \qquad (9)$$

The usefulness of network coding in coupon collection is seen by comparing this result (eq. 9) with the previous one (eq. 4) for classical coupon collection problem. The value of $E[N_S]$ can be made close to S by increasing the field size q. This can be achieved at the price of increased computation complexity for solving a system of linear equations over a large field size. If the field size is designed carefully, there is a possibility of complexity reduction.

## 5. Proposed system:

### 5.1 System model

In this article, node capacitated P2P network model is used rather than edge capacitated network model [25]. Each node's uploading and downloading rate is upper bounded. This is a suitable model for P2P networks since bottlenecks normally are at the edge of the access networks rather than in the core of the Internet. Hence, in P2P networks, capacity constraints are put on peers on the Internet instead of links between peers. Because of an experimental observation that residential broadband links have asymmetric upload and download rates, some existing research work further imagines that there is no restriction on the downloading rate for each node [26]–[27]. This assumption is not considered because in practice each peer in the network has only a restricted download bandwidth. Further the following assumptions are made to model the P2P network to analyze the performance of proposed P2P file sharing mechanisms

1. The original content is divided into segments and further segments are divided into smaller units of messages called as chunks. The number of chunks in each segment should be less than few hundred with the chunk size less than a few megabytes, in order to make sure reasonable encoding and decoding complexity. The optimal chunk size is 2 − 32 KB, which offers fastest encoding and decoding. This optimal value increases with the number of chunks in each segment.
2. There is a single node which has the original content and it has to be distributed to n participating nodes in the network. Out of those n nodes some of the nodes are inside the campus network and remaining nodes are located outside the network. All are interested to download the same file from the single source or server.
3. In our model, all peers within a campus network are organized into several groups by using the similarity interest group algorithm (SIG) after reception of some set of native packets of original content. Each group is headed by at least one super-peer. All super-peers in the network forms an upper overlay named the super-peer layer.
4. The super peer sends linear combinations of the original n packets in a segment, while peers within a group recursively and equally combine their collected packets at random and generate new coded packets that they disseminate through the group. Hence, each coded packet holds a linear combination of the original packets. This linear combination is described by the coding vector of size n symbols over Galois Field GF($2^q$), that states which linear combination each coded packet holds.

In the Fig. 2., the proposed P2P model is described in which peers are grouped into three different groups and each group headed by one super peer. The super peers are connected among themselves.

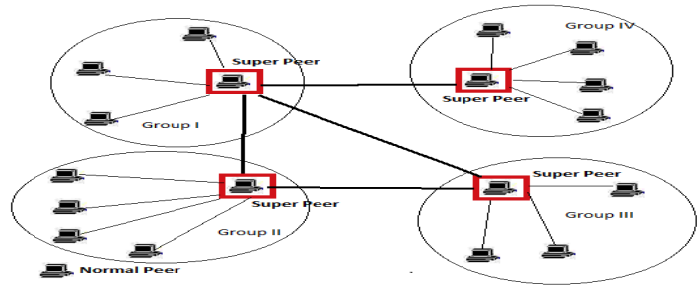

Fig. 2. Proposed System Model

The proposed system consists of two stages. In stage one, formation of P2P similarity interest group with different content of same file is carried out. Dynamic coding is carried out in the stage two of the proposed system.

### 5.2. Stage I - Formation of P2P similarity interest group with different content:

The peers can form groups according to their content interest by using Similarity Interest Group (SIG) algorithm after the native packet transmission phase from the server. A P2P similarity interest group is a set of irregularly connected nodes with each node providing its resource such as content, bandwidth and storage to the rest of the nodes in the group but those members are having dissimilar content. The number of groups created based on the number of nodes currently interested to download the content. Each group is headed by one super peer who has full control over all other nodes in the group. When a node in the group downloads content, it attempts to retrieve the content from the other nodes in the same group. Here the data exchange among the nodes in the group is coded packet. As an example of a P2P similarity interest group, consider a campus network or an ISP network or similar kind of IoT based network [28]. As shown in Fig. 3, the peers in a campus network are high speed links, and the high speed LAN is linked to the global network through a lower speed access link. If the peers reside within the campus network do not organize themselves as a P2P similarity group, they may independently retrieve the same content from peers outside the campus. It leads to the congestion in access links and wasting the resources. The campus network could make more efficient use of its resources if the peers are organized as a P2P interest group. In a P2P interest group, the peers in the campus network would cooperatively maintain a managed number of content, and would attempt to retrieve files internally before retrieving them from outside the campus network [29-30]. Note that the main goal of forming the interest group is to minimize traffic on the network links as well as minimize the coding complexity and retain the benefit of network coding fully.

In our model, there is an assumption that each peer attains content interest from other peers on the network, where $C_j$ is a function of the probability that the other peer will respond to the peer's content interest requires. This kind of measures are operational using $D_i$ (Dissimilar content with Similar Interest) such that the content interest of a peer $P_k$ obtains from another peer $P_l$ is given by,

$$C_j (P_k, P_l) = D_i(P_k, P_l) \times K(P_l) \qquad (10)$$

where $K(P_l)$ is the amount of content contributed by $P_j$

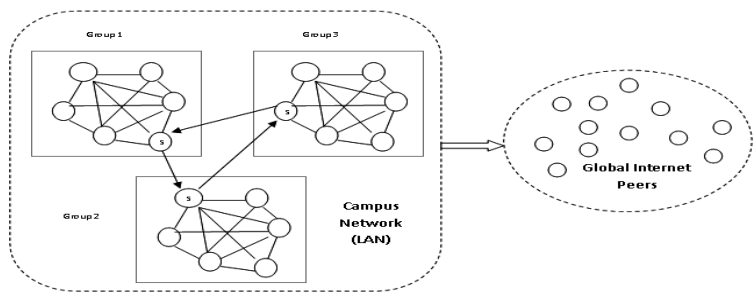

Fig. 3. Formation of P2P Group

---

**Algorithm 1- Similarity Interest Group with different content**

**Input:** Collection of Nodes, $\sum N_j$ where $0 \leq i \leq p$, p is the total number of peers in the network.

**Output:** Group of Peers based on their content interest
1. Begin
2.    For each nodes $N_i$ in the campus network
3.      Sends greeting message to its neighbor Nodes $N_k$, message includes the content it has
4.      If ($C_i (N_i, N_k)$ = true)
5.        $N_i$ replies with attach message
6.      Else
7.        $N_i$ replies with refuse message
8.      End if
9.    End loop
10. End

---

**5.3. Stage II – Dynamic coding stage**

The second stage of the proposed system consists of two phases such as native packet transmission phase (NPTP) and coded packet transmission phase (CPTP). The NPTP happens before the group formation for certain duration. During the NPTP phase, the n on coded packets or native packets are transmitted, in which the source simply sends a fixed number of original packets one by one. Initially first n transmissions are NPTP. The value of n determined based on the size of the content. When the content size increases, the value of n also increases. Both are directly propositional.

The CPTP phase is performed after the group formation among the peers in the network. During the CPTP phase, instead of using a complex mechanism to locate the optimal set of packets for XOR-ing like the previous schemes do, here the simple approach to group packets into different groups is used. Then, the source deals the packet group by group. Only when all nodes have recovered all the packets of the current group, the source will keep on to send the next set. During the coded transmission phase of each set of packets represented by $C_p$, after a necessary parameter initialization

(step 2), novel mechanism to determine the proper different linear combination of the coded packets is used. After the transmission of these coded packets, if some node cannot recover all packets $C_p$, the Static mechanism will update some parameters (Step 4) and then continue to obtain a new coding packet (Step 3) to send. In the following, the main step of the coded transmission phase of the proposed system is described.

### 5.3.1. Procedure 1: Packet Set Formation

If suppose $N_n$ naïve packets need to be sent in the current segment, then $N_n$ packets are grouped into $N_n / C + 1$ group, such that group has the same one to one correspondence C and last group has the one to one correspondence ($N_p$ mod C). For the last group with one to one correspondence ($N_p$ mod C), add additional C - ($N_p$ mod C) packets with only bits zero into this group, and also assign all indicators $L_{i,j}$ of these additional packets to zero. For a set of native packets P = $\{P_1,....,P_n\}$ and one its coded packet $P_c = \Sigma\ c_i p_i$ for the limit i-1 to n over a finite field $F(2^q)$. We call $(c_1,...,c_n)$ for $P_c$'s coding vector over P, and denote it by $(P_c, P)$. Hence, the major problem now is to select the coding vectors for generating each coded packet. Before transmitting each set of coded packets, the coding node needs to perform the constraint initialization.

### 5.3.2. Procedure 2 (Constraint Initialization)

For a given group of packets $G_p$, let $N_n^{p,i}$ is the number of packets $G_p$ of native packets that node $R_i$ has not received yet. Assign the values of $N_n^{p,i}$ by $N_n^{p,i} = \sum_{i=1}^{n} L_{i,j}$ and also assign the set of coding vectors C as C = $C_{n,q} \setminus \{(0^{i-1}, 1, 0^{n-i})\colon i \in \{1,.., n\}$ as initial value and $i^{th}$ native packet in $G_p$ has been obtained by at least one node, where $C_{n,q}$ is the maximum set of n-dimensional vectors over finite field $F(2^q)$, which holds n distinct unit vectors (1,0,0,…,0),…, (0,0,…, 1) and any n vectors of it are not linearly dependent. The objective of this step is to generate the linear independent coded packet with minimal coding density. After the constraint initialization, now each node can select the optimal coding vector for each transmission during the each coded transmission.

### 5.3.3. Procedure 3 (Vector Selection)

Randomly choose a coding vector c in C and allow C $\leftarrow$ C \ {c}. Then let vector c is a coding vector over $G_p$ to generate a coded packet. The received packet either native or coded packet is called as non innovative for the received node if this packet is on hand or can be obtained by linear combination of its earlier received packets. Hence, the node $R_i$ needs to collect at least $N_n^{p,i}$ innovative packets to reconstruct all its packets in $G_p$. During the code packet transmission, a node $R_i$ that has not received $N_n^{p,i}$ innovative packet is called as backlogged. For each backlogged node, the vector selection in procedure 3 is independent of the coding vectors of its previously received packets, such as resulting coding vector is innovative to it. This approach reduces the expected number of transmissions required for the delivery of all native packets in $G_p$. After transmitting a coded packet, the source needs to update the constraint $N_n^{p,i}$ and C as follows to the response from the nodes.

### 5.3.4. Procedure 4 (Constraint update)

For each backlogged node $R_i$ (with $N_n^{p,i} \geq 1$), if it properly receives $P_k$, $N_n^{p,i} \leftarrow N_n^{p,i} - 1$. For each coded packet $P_c$ of $G_p \cup$ {transmitted coded packets from Gp}, if $\Sigma L = 0$ with subscription for L is i,j and limit for summation is between n and $N_n^{p,i} >=1$, then the encoding vector of $P_k$ can be used again and thus C $\leftarrow$ C U { $E(P_k,G_p)$ }.
This coding scheme is formally described as follows

| Algorithm 2 - Constraint Update |
|---|
| 1. Begin |
| 2.     Send N native packets one by one and create the vector table |
| 3.     Perform packet group formation to group $N_n$ native packets into C group. |
| 4.     For i=1 to C do |
| 5.         Let $G_p$ is the I th group of native packets |
| 6.         Conduct constraint initialization to initialize parameters $N_n^{p,i}$ and C |
| 7.         Until exist one or more backlogged peers such as $\exists i\ N_n^{p,i} > 0$ do |
| 8.             Perform vector selection to select encoding vector and generate coded packet P |
| 9.             Repeatedly transmit coded packet P until at least one backlogged peer receives it |
| 10.        Perform constraint updating to keep parameters updated one |
| 11.     End while |
| 12.     End for |
| 13. End |

### 5.3.5. Data Dissemination:

The content is divided into several blocks with each block represented by a symbol in $GF(2^q)$. Before transmitting the content, the server has to code the content. The encoding is over a Galois field $GF(2^q)$. The first set of blocks is coded and then the second set of blocks is coded, and so on. Assuming the field size is greater than the number of messages to be transmitted in each group. Each block is encoded into a different message using the coding scheme described in above. Hence, any messages of these set of messages can be utilized to decode the original blocks.

### 6. Simulation and performance evaluation:

Evaluating a new mechanism and protocol in real surroundings, particularly in its early stages of development, is not possible. PeerSim simulator is used to simulate and evaluate the performance of the proposed mechanisms. It has been developed with extreme scalability and support for dynamicity in mind.

The following system is used to evaluate the performance of the proposed system

1. **Traditional non network coding mechanism (TNNC):** It is similar to the BitTorrent like system where the server divides content into different chunks and transmits to the nodes in the network. Then each node needs to collaborate with each other to download the entire content.
2. **Flat network coding mechanism (FNCM):** It is similar to the Avalanche like system where each peer exchange coded packet among themselves instead of native packets. In this system, P2P distribution based on exchanges of individual blocks is performed in a decentralized fashion.

The proposed system is compared against the existing system using the following metrics

**Throughput**: It is defined as the number of bytes uploaded by all the peers in the network per second. This measures the speed at which content is disseminated.

**Average finish time:** It is defined as the time to complete the downloading of the entire content by all the peers in the network.

**Maximum finish time**: While the average finish time reveals the overall system performance, the maximum evident the worst case. It is defined as the maximum amount of time peer takes to download the entire content.

**Failure rate:** It is defined as the number of peers unable to finish its downloading due to peer churn. All these metrics depend on the behavior of the peers, network environment, incentive mechanisms, etc.

**Link Stress**: It is defined as the ratio between the number of total packets sent over a link and the number of distinctive packets transmitted over the link. For example, a stress of 3 corresponds to case where each packet is transmitted thrice over a link. It is used to evaluate the effective use of bandwidth available for peers to exchange the content.

The performance of the proposed system is analyzed in two different arrangements:

(i) **Homogenous Arrangement**: In this arrangement, peers have the same link capacities, and overlay networks are formed randomly. The content is sent after the overlay network and grouping is formed.

(ii) **Heterogeneous and Dynamic Arrangement**: In this arrangement, peers have the different link capacities, and overlay connections are created randomly. Peers join the system during the file transmission and stay in or leave the system after downloading the entire file.

### 6.1. Homogenous arrangements

In the homogeneous arrangement, peers have equal link capacities [31]. The simulation is divided into two steps. First, overlay construction period and the group formation: A number of nodes are selected randomly and connect to the system in sequence. Second, content transmission period: The server transmits the content to the overlay network. The average finish time curves of the proposed system, TNNC and FNCM are plotted in Fig. 4. The proposed system is simulated with different values as shown in the fig.. The average download time of nodes is arranged in an ascending order. It can be seen that the average download time of the proposed system is 20 - 25% shorter than that of the existing system. We observe that the download times of TNNC spans a broader range, which leads to a larger variation than that of FNCM and the proposed system. This is for the reason that in TNNC, multicast tree is constructed and the two peers with the largest difference in download time are a child of the origin and a leaf, respectively [32]. This difference may be further very huge depending on the overlay topology. On the converse, FNCM and our scheme build a mesh to disseminate the content. As a result, the distance between the highest level peer and the lowest level peer is reduced. The reason for the proposed system performs better than FNCM with respect to throughput is two folds. First, FNCM implements random network coding which may produce some non innovative messages. This will be the reason for unnecessary delay for some nodes to carry out decoding. The proposed system implements a deterministic network coding scheme with low complexity and control overhead to promise that each message is the innovative one. Second, although proposed system requires some topology restrictions to the overlay network because of deterministic network coding, the restrictions are flexible and resilient.

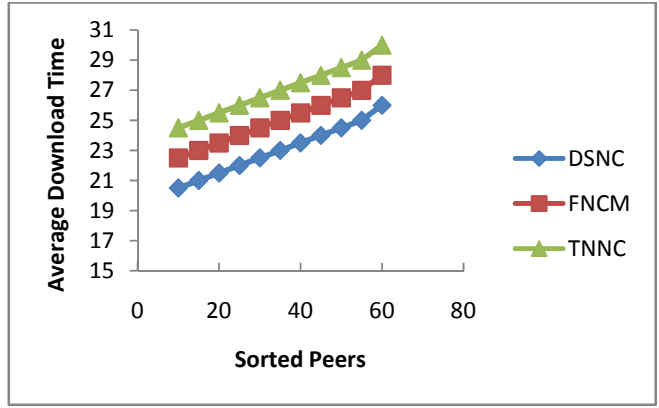

Fig. 4 . Average Download Time in homogeneous Arrangements

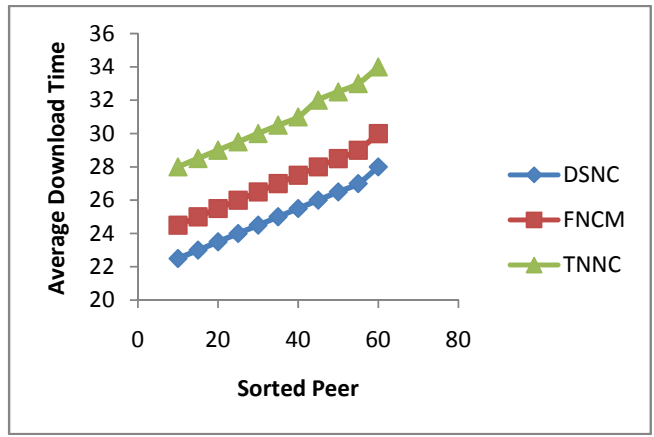

Fig. 5 . Average Download Time in homogeneous Arrangements with link failures

Average download time is calculated by considering the time (in seconds) needed for each peer in the network to download the entire content in each run of the system with different number of sorted peers. It is represented in Fig 5, 6, 7 and 8.  Fig. 5 shows the average download time in the presence of some of the physical link failures. It is assumed that the probabilities of link failure on the different physical links are independent. It is observed that the average download time of DSNC and FNCM are much shorter than that of TNNC. With the same link failure, the average download time of DSNC is slightly shorter than that of FNCM. With referred to the previous simulation result without any link failure, the increase of average download time is shorter for FNCM and larger for TNNC. The throughput reduction is small for both DSNC and FNCM when physical links are not stable.

The link stresses of the proposed system, FNCM and TNNC is evaluated and it is shown in Fig. 6. When the number of peers in the network is small, all the three mechanisms have similar link stress. However, when the number of peers is more than 100, the links stress of DSNC is less than other two mechanisms. The reason for this is similar to that for the larger average download time variance of FNCM.

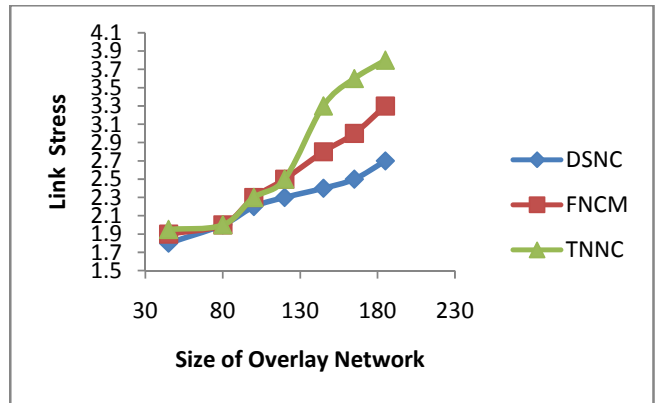

Fig. 6. Link stress of the Homogeneous Arrangements

### 6.2. Heterogeneous and Dynamic arrangements

In this arrangement, peers have the different link capacities, and overlay connections are created randomly. Peers connect to the system during the file transmission and stay in or leave the system after downloading the entire file.

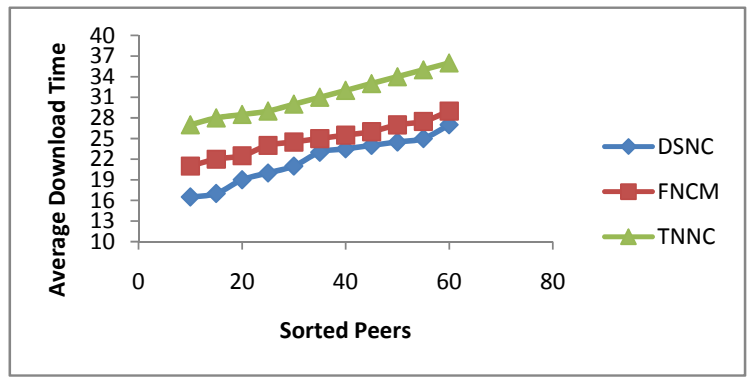

Fig. 7. Average Download time in dynamic arrangements - Peers stay in the system after downloading the content.

Fig. 7 shows the average download time comparison when peers stay in the system after receiving the content. It is observed that the average download time of all three schemes increases with compared to the homogeneous arrangements. For DSNC and FNCM, this is because of lack of peers which assist forwarding the content. For TNNC, this is because of the join delay of peers in the network. The increase of average download time for TNNC is more than that of DSNC or FNCM, which shows that P2P network achieve better resilience to dynamic joins than tree-based approaches. The peers with different download times are not evenly distributed as in the homogeneous arrangements. The largest average download time for FNCM is nearer to that of DSNC. This is because of DSNC requires to download the content from peers in different groups while FNCM has no such conditions.

Fig. 8 shows the average download time comparison when peers leave the system after receiving the content. It is seen that the average download times of DSNC and FNCM are increased by approximately 15%, while that of TNNC is increased by about 50%. Both DSNC and FNCM have similar resilience under the peer's churn. However, DSNC achieves a little higher throughput. It shows that DSNC attains great resilience under dynamic peers join and leave the network. Even though DSNC implements deterministic network coding, the overlay topology in DSNC is relatively flexible. In addition, by adding redundant links, the resilience can be enhanced dramatically.

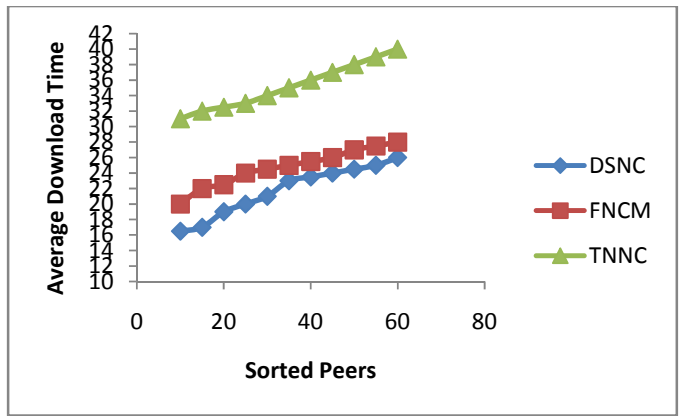

Fig. 8. Average Download time in dynamic arrangements - Peers leave the system after downloading the content

**6.3. File downloading progress and Additional message overhead**

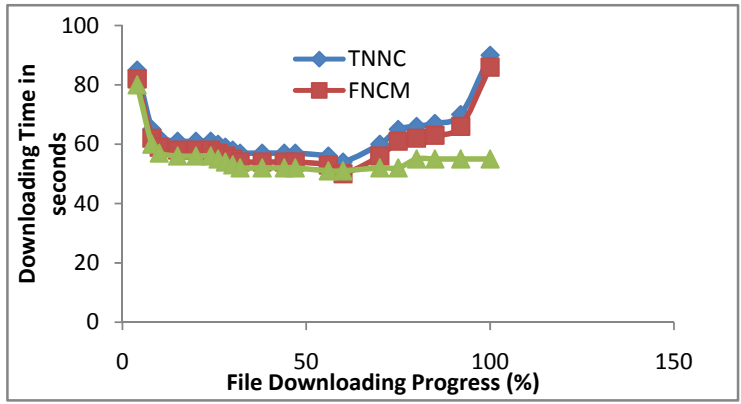

Fig. 9. File Downloading Progress

Fig. 9 shows the ratios of download times for each segment against the total time needed to download entire content. In Fig. 9, under circumstances where network coding techniques were not utilized, more time was needed at the early and end of time. It is found that only a few peers had the original content at the beginning, and many peers demand chunks for download from these few peers. The time needed to download chunks from these peers will be delayed because of a limited uploading capacity. This incident also happens in schemes that use network coding. On the other hand, in the final phase when content is almost completed, a great deal of time is needed to complete file downloads due to the difficulty of finding a few remaining pieces. Fig. 9 also shows that for schemes that utilize the network coding, the proportion of download time comes down significantly when the content approaches download completion. But, the problem of coded packet coefficient linear dependence still occurs in the final stage, requiring a greater amount of time to complete downloads. By contrast, the additional ratio increase is somewhat smaller than in schemes without network coding. Fig. 9 reveals that dynamic smart network coding with group formation schemes requires spending approximately 50 % of downloading time on the final stage of the content transmission. It can be seen that the proposed network coding schemes still face the problem of large piece linear dependence in the final stage.

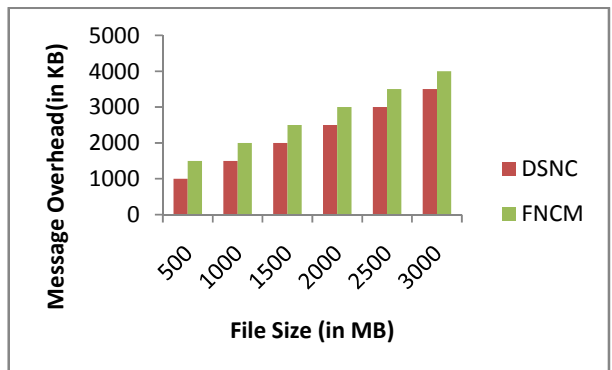

Fig. 10. Additional message overhead

The additional required message overhead in coded packet is shown in Fig. 10. The additional message overhead produced look to be notably greater than in mechanisms which do not use network coding. In mechanism which did not employ network coding, the TNNC scheme generated a message overhead of only 264 KB for 122 MB content. The scheme which used network coding (FNCM) needed 1,030 KB of extra overhead. But in case of the proposed mechanism require less than 1,030 KB since the number chunks are reduced in each generation. It reduces the length of the message overhead carried in each coded packet with compare to FNCM. The Fig. 10 shows comparison between FCNM and DSNC only since TNNC does not require any additional message overhead.

## 7. Conclusion

In this paper, the coupon collection problem related to P2P network and the effectiveness of network coding for P2P file sharing with mathematical analysis are investigated. Based on the study, the comprehensive P2P file sharing scheme based on network coding such as Dynamic Smart Network Coding (DSNC) technique to further enhance the performance of network coding in P2P content distribution with respect to scalability, reliability and resilience against peer's churn is proposed. With compared to other available network coding scheme, the proposed coding scheme run in polynomial time. Since peers are grouped into different groups and chunks are coded and distributed based on the group it belongs, coding density and complexity of carrying out the coding and decoding operations are minimized. It also minimizes the control information overhead (coding vector) in the coded packet. Additionally, the simulation results reveal that, compared with the other content sharing schemes, the proposed group based approach can more effectively reduce the bandwidth consumption in access link, decrease link stress and enhance the throughput. Our future research work is to focus on how to optimize the system under network congestion and optimize the process of group formation and selection of super peers.